\documentclass{article}
\usepackage{graphicx} 
\usepackage{xcolor}
\usepackage{natbib}

\title{A UDP Packet Format Establishing Adress Event Representation Communication Between Remote Neuromorphic and Biological Setups}
\author{CG Mayr$^{1,*}$, RM George$^{1,*}$, M Tambaro$^{2}$, G Indiveri$^{3}$}
\date{September 2024}

\begin{document}

\maketitle

1) TU Dresden

2)  Università degli Studi di Milano-Bicocca

3) INI Zurich

*) The first two authors contributed equally

\begin{abstract}
In the field of brain-machine interfaces, biohybrids offer an interesting new perspective, as in them, the technological side acts like a closed-loop extension or real counterpart of biological tissue, instead of the usual open loop approaches in tranditional BMI. To achieve a credible counterpart to biological tissue, biohybrids usually employ one or several neuromorphic components as the hardware half of the biohybrid. However, advanced neuromorphic circuit such as memristor crossbars usually operate best in a dedicated lab with corresponding support equipment. The same is true for biological tissue, which makes co-locating all of the parts of a biohybrid in the same lab challenging. Here, we present as solution to this co-location issue a simple method to connect biohybrids via the internet by a custom UDP packet format. We show that the characteristics achieved with our solution (jitter, delay, packet loss, packet reordering) on a standard internet connection are compatible with various biohybrid processing paradigms, and we present a short three-ways experiment as proof-of-concept. The described UDP format has been employed to link biohybrids and neuromorphic circuits in four different EC-funded projects.
\end{abstract}


\section{Introduction}
\label{sec_intro}
Biohybrids are systems consisting of a technical and a neurobiological component, where both are linked very closely in a mutual overall system integration. Biohybrid belong to the wider field of brain-machine interfaces (BMI), but are somewhat distinct in that the aim is not primarily the interfacing for a certain application such as a neural prosthesis. Rather, they are targeted at deriving an overall system that blurs the boundaries between the biological and the technological side, with the technological side enabling to understand, gently control and virtually extend the biological part. Seamless dynamical integration of hardware and biology makes such a hybrid system most effective, where we define seamless as that the hardware neural network operates in the same dynamical regime as its biological counterpart, and tight coupling of both generates a meaningful joint dynamics \citep{keren2018closed}. Thus a biohybrid usually establishes closed-loop bidirectional interaction between the two sides (as opposed to the usual single direction for conventional BMI) and has an overall processing function that is spread out over both components and uses them synergetically.
As the technological component of a biohybrid, a neuromorphic circuit is usually chosen. Neuromorphic circuit design is a branch of semiconductor circuit design that implements circuits which mimic aspects of the behaviour of living nerve tissue. Popular aspects to be implemented are e.g. pulsing communication between neurons and biological dynamics/behaviour in neurons such as regular spiking, accommodating, bursting, etc. Plasticity is also implemented often, e.g. in neurons as spike frequency adaptation or in synapses as facilitation/depression or as longer-term learning such as spike-time-dependent plasticity \citep{george2020plasticity}.

The usual interface between the biological and circuit components in biohybrids is an event-coded stream, called address-event-representation (AER)  \citep{boahen2000point}. On the neuromorphic circuit side, AER events encode neuron spikes. On the tissue side, AER events encode for some aspect of neurobiological signals, e.g. a spike identified via spike sorting from a microelectrode array, or a delta-encoding of extracellular measurements \citep{cartiglia20234096}. This representation is sparser and therefore more efficient than the conventional AD-conversion of neurobiological signals carried out in most BMI, and it is thought that a biohybrid is better able to integrate both sides into a cohesive processing unit if information encoding is done by the natural exchange language between neurons, i.e. spikes. By encoding culture output as AER events, neuromorphic systems can then process these events using hardware emulations of SNNs with complex dynamics and on-line learning abilities. 

In the last couple of years, experimental neuromorphic technologies such as memristors have been increasingly integrated in biohybrids \citep{tzouvadaki2023interfacing,mikhaylov2020neurohybrid}. However, this creates some unique problems, as few labs are equipped to support several complex, disparate and partially experimental technologies such as analog neuromorphics, biological cell cultures and memristors at the same time. Rather, each systems works best in its home lab with its dedicated support infrastructure. Thus, we have come to the conclusion that a communication method between remote sites was required that would allow interfacing of neuromorphic and biohybrid components distributed across labs, and that uses event encoding. 

For this purpose, a tentative standard was derived in the Capo Caccia neuromorphic engineering workshop, and subsequently refined based on the use experience in several neuromorphic and biohybrid EC and national projects. As stated, the interface is primarily event-based, with extensions reserved for e.g. additional non-binary (i.e. non-spike) data. An Ethernet communication format was established that employs the user datagram protocol (UDP) \citep{postel1980rfc0768}. UDP as a simplified Ethernet packet allows for a relatively lightweight hardware or software implementation and low-latency communication, while being widely supported by standard network hardware and software. In this paper, we first give a more detailed account of the original AER standard (Sec. \ref{sec_AER_intro}), plus its modification in current neuromorphic systems. We then describe the UDP packet encoding as required by the biohybrid constraints (Sec. \ref{sec_UDP_AER}). The threeways biohybrid remote setup used in several projects is illustrated (Sec. \ref{sec_threeways_setup_intro}), and requirements for the UDP transmission characteristics are derived from e.g. timing constraints set by plasticity experiments (Sec. \ref{sec_plasticity_timing_requirements}). 
In the results section, we first show that this UDP format and its practical characteristics on a standard internet connection are compatible with various biohybrid use cases (Sec. \ref{sec_results_bidirectional}). Lastly, a proof-of-concept is detailed, showing a practical biohybrid consisting of components in three different labs being interconnected by our UDP communication (Sec. \ref{sec_results_threeways}).

\section{Methods}
\subsection{AER and related event-based packet communication}
\label{sec_AER_intro}

Adress Event Representation (AER) is traditionally a format to transmit spike data to/from a neuromorphic chip, especially the identity of the neuron emitting a spike. This information is transmitted asynchronously, i.e. without a time stamp, to be processed at the receiver as it comes in, in real time.
It was first proposed by Mahowald and Sivilotti for a neuromorphic vision system, and has since been extended and standardized in various directions (see \citep{boahen2000point} for a review and history). It is essentially a time-division multiplex communication link, where the speed difference between electric communication links (MHz-GHz) and biological axons (<1 kHz) is exploited to communicate multiple virtual axons across a single electrical communication channel. As mentioned, compared to conventional synchronous time division multiplex, AER transmits asynchronously, to take full advantage of the real-time capability of neuromorphic systems, where time essentially models itself and information/events are transmitted as soon as they are generated. Specifically in the context of event-based sensor systems, AER has the advantage to thus keep a very fine time resolution of when a certain sensor signal has happenend \citep{yang2015dynamic}.  

\begin{figure}[ht!]
\centering
\includegraphics[width=110mm]{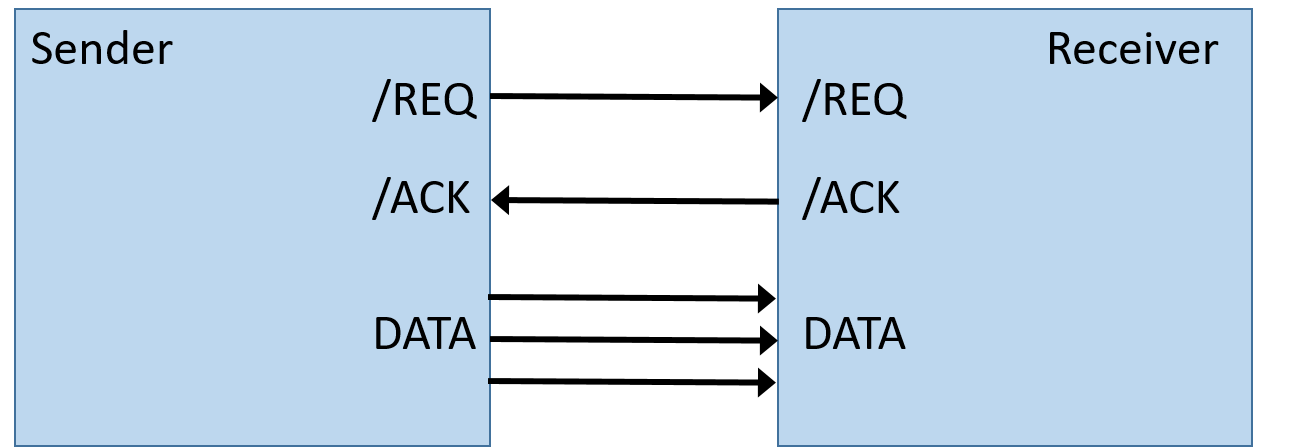}
\caption{Sender and receiver of AER link, signals}
\label{fig_AER_link}
\end{figure}

Fig. \ref{fig_AER_link} Shows the structure of a unidirectional, single sender, single receiver parallel AER link \citep{boahen2000point}. It is comprised of a inverted-logic request line from sender to receiver, an inverted-logic acknowledge line and the data lines. Commonly, 16 data lines are used, so a total of 65k spike sources can be encoded. Usually, the sender contains an arbiter to handle concurrent spikes in some kind of queue logic, either in 1D (for a neuron row) or in 2D (for a pixel array or similar) \citep{boahen2000point}. 

In the idle state, the sender drives /REQ to logic 1, the receiver drives /ACK to logic 1, and the data lines are considered invalid. To begin a transaction, the sender first drives valid
signals on the data wires, and then drives /REQ to logic 0. In response, the receiver senses this valid data; this sensing operation may take an arbitrarily long amount of time. Once the receiver has sensed the data, it drives /ACK to logic 0; by this action, the receiver has removed the requirement of the sender to supply valid data, and data sending can be released. Upon sensing /ACK is at logic 0, the sender must drive /REQ to /ACK logic 1. Upon sensing /REQ is at logic 1, the receiver must drive /ACK to a logic 1;  Once the sender has sensed /ACK is at logic 1, i.e. the handshake is fully executed, it is free to start a new transaction. With AER, fairly large multi-chip setups have been constructed, for example in the Caviar project \citep{serrano2009caviar}. To facilitate adoption to multiple chips, automated methods have been derived to describe and synthesize the required logic blocks \citep{mostafa2013automated}. There have also been some efforts in the past to extend the AER protocol with other information formats such as configuration \citep{Qiao_2015}, as AER by its nature simply transmits digital packets in an asynchronous fashion, i.e. AER is not restricted to spike data. 

While the above parallel AER interface with handshake above has been quite sucessful, it is limited like conventional parallel full swing interfaces to the kEvent/s to MEvent/s range, and to a physical point-to-point connection with a fairly large cable. Thus, to make routing of events more flexibel, reduce mechanical overhead and increase speed, various derivatives have been implemented. For instance, \cite{fasnacht2008serial} have derived a serial version at reduced swing to increase speed. Another serial, reduced swing versions of AER has been proposed for the Facets/BrainScales waferscale system \citep{hartmann10}, which needs communication in the GEvents/s range due to its accelerated time operation. Interestingly, the Brainscales communication, as well as spike communication of the SpiNNaker system \cite{plana2020spinnlink}, have changed to transmitting packets of spike events on these links, instead of just the pure data exchange by handshake, as in Fig. \label{fig_AER_link}. These packets are sent of in a fire-and-forget manner, i.e. without a handshake to establish that the spike arrived at the receiver, as the roundtrip delay would be too long and reduce available bandwidth too much. In order for such a packet-based, unidirectional link to operate in the same way as the handshaked, bi-directional original AER, such a neuromorphic communication system has to fulfil five constraints: (1) Packet loss needs to be fairly low and acceptable for the particular neuromorphic benchmark running, (2) to avoid queues filling in the communication path and to keep real time operation, receivers (aka neurons) need to be able to integrate and/or process incoming packets within a short time frame, e.g. a millisecond, (3) delay along the whole communication pathway also needs to be within a fairly short time so that spike packets arrive in real time and (4) routing/network jitter needs to be  compatible with the neuromorphic use case, and (5) as a packet-based network opposed to the original AER can loose packet order, meaning spikes can arrive in a different order than in the one which they were sent, the neuromorphic use case should be tolerant of the level of packet order loss encountered in a particular setup. 

\subsection{The UDP AER protocol}
\label{sec_UDP_AER}

To solve the problem mentioned in the introduction, i.e. link diverse setups in different labs into one functioning biohybrid, the natural choice is of course an internet-based connection. By establishing a custom AER format carried in an internet packet, the various setups (memristor, cell culture, neuromorphic setup) can be kept independent, separate packet sending and receiving backends are integrated into each setup that convert internal AER or other representations into the custom packet format, and standard internet hardware (routers, switches, etc) can be used to establish the overall biohybrid. 

From the wide variety of packet and communication standards supported by internet backbones \citep{kessler2004overview}, we have chosen the UDP packet format, a simplified Ethernet packet \citep{postel1980rfc0768}. UDP allows for a relatively lightweight and low-latency hardware or software implementation and low-latency communication, while being widely supported by standard network hardware and software. Typical UDP transmission latencies and jitter are also compatible with the timing required for biological setups, as shown later in results.
  
\begin{figure}[h]
\centering
\includegraphics[width=1\textwidth]{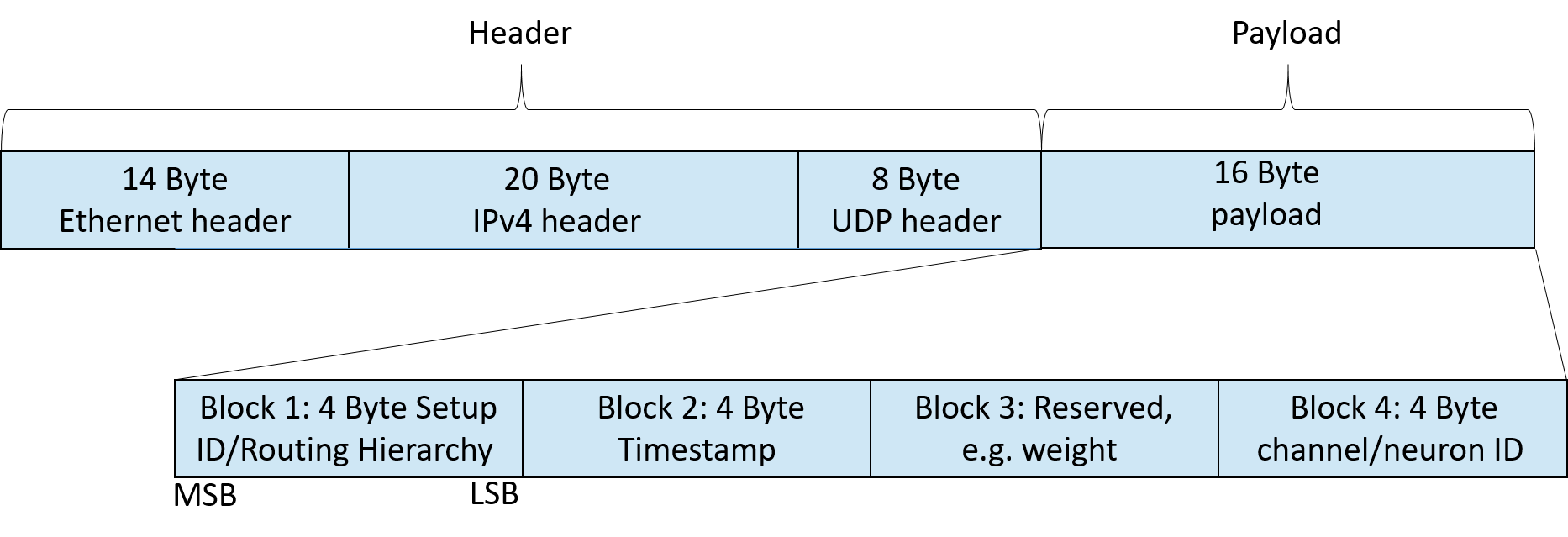}
\caption{UDP packet format for event transmission in hybrid setups, with its four information containers (called block 1-4)}
\label{fig_UDP_format}
\end{figure}

 Fig. \ref{fig_UDP_format} illustrates the 58~Byte UDP packet format we derived for biohybrid communication. The outermost wrapping is the standard 12~Byte Ethernet header (layer 2 of the OSI model) \citep{alani2014guide}. The next one is the 20~Byte IPv4 header of layer 3 in the OSI model, containing routing and packet type information. The innermost wrapping is the 8~Byte UDP header, containing e.g. a checksum that allows us to check for corrupted event data packets, and the length information of the UDP data field. Theoretically, the 16~Bit length field can encode a maximum of 65~kBytes data field, but in practice, UDP is delivered with much smaller payload for faster routing, as is also the case for our packets, as we define an 16~Byte/128~Bit data payload. 

The payload consists of four 32~Bit containers, called blocks. The first 4~Byte/32~Bit contain hierarchical routing information for larger systems, e.g. for the target system that decodes the spike target address. In the specific use case outlined in Sec. \ref{sec_threeways_setup_intro}, this ID is used to code the information about the setup the packet is coming from. This is necessary as the CU node in Dresden routes events to/from multiple other setups, needing a method to disambiguate packets. Thus, this field also acts as an identifier, allowing the interpretation of an arriving package independent of the possibly changing IP addresses of the sender, and instructs the receiver on how to process incoming data. 

The second 32~Bit encode source timestamp. The default equivalent of the LSB timestamp is $50~\mu s$. This value was chosen since we cannot go much lower due to latency of Ethernet switches, even for a local network \cite{prytz2005real}. On the other hand, much larger values could lead to timing precision problems respectively mask the sub-millisecond timing resolutions available on modern MEAs \citep{pirog2018multimed}. With a 32 bit timestamp, we can thus have a ca. 60~h experiment before the timestamp wraps, easily satisfying most experimental paradigms. If track is kept of the wraparounds, this time can even be arbitrarily extended, while still unambiguously identifying every pulse for any reasonable experiment (i.e. where the max distance between two consecutive pulses from the same source is lower than 60~h). Transmitting the time stamp also enables us to check for out-of-sequence packets at the target, and we can correct jitter along the transmission path by delaying packets, if that additional delay is compatible with the information processing paradigm of a particular biohybrid experiment. 

Block 3 is reserved for custom payloads. These custom payloads could be amplitude values in an interface to MEA, weight value of a synapse (i.e. to enable graded spikes), or additional information such as firing rates. Block 4, i.e. the last 32~Bit, contain pulse source ID, equivalent to 4 billion possible spike sources. This may be a bit of overkill, but in terms of payload to header ratio, the payload is still small, and keeping the blocks large, they can also be substructured in the future, i.e. adapted for a variety of usecases. 

There are several protocols for streaming data implemented based on UDP packets, such as the real time transport protocol (RTP) \cite{rajreal}. The reason we did not directly use one of those is that they contain a fair amount of overhead for data stream assembly, have complex stream encodings that serve as packet-loss compensation mechanisms, or were not flexible enough for our purposes. However, at this reduced overhead, we still implement the basic elements of RTP, i.e. payload characteristics identification, source identification, and time stamping \cite{rajreal}. We did not implement sequence numbering, as our measurements show that packets mostly stay in order, and we can use the time stamps for out of sequence packets. As our measurements also show the packets are not overly lost, and jitter is not too high, we are compatible with requirements of biohybrids without a lot of that overhead, making implementation on various platforms significantly easier. 

We can detect defect UDP packets via the standard UDP CRC check, so we can drop those with false data. There is no mechanism for handshaking or resending of packets, but with the measured small packet losses, this was not deemed necessary. As shown in the next section, we can establish an absolute time base based on the timestamps, which enables compensation of both delay and jitter to a certain extent, and incidentally also takes care of reordered packets. One mechanism that usual streaming protocols contain which we really missed in our biohybrid lab sessions was some way to scan through open ports in firewalls in order to tunnel our information through them. In standard UDP packets like our format above, the port is hard coded in the packet, so we had to manually work through ports. Also, our custom code usually was not signed/certified in any way a firewall would recognize as a legitimate data stream. Thus, part of our initial work in each lab was to establish permanent routes and ports for our event packets through the lab firewalls with the local IT technicians.    

Thus, the general approach was that simplicity is key in order to be able to use it in various heterogeneous setups: memristor, analog neuromorphic, digital neuromorphic, biological. Across the various projects, implementations were carried out in (1) Python for a memristor setup at University Southampton, (2) C++ for a neuromorphic setup at INI Zurich and a biological setup in Padova, (3) a hybrid Matlab/C++ version that interfaces to an xPC real time system running a biological setup at Technion Haifa, and (4) a Verilog version for the FPGA backbone of a neuromorphic setup at TU Dresden.  

\subsection{The fourways biohybrid setup}
\label{sec_threeways_setup_intro}

\begin{figure}[h]
\centering
\includegraphics[width=1\textwidth]{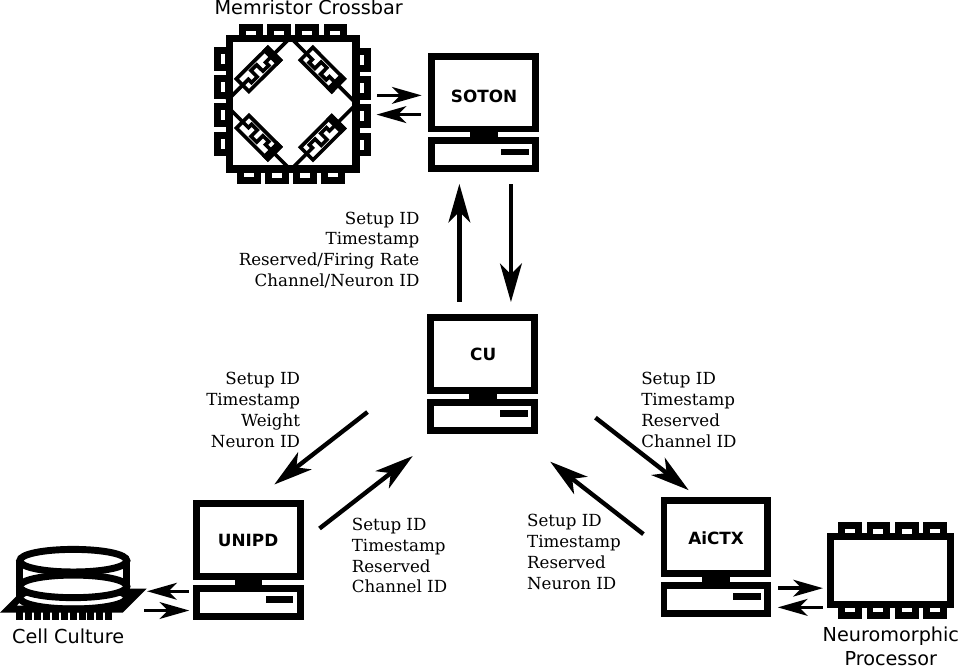}
\caption{Star-shaped topology of proof-of-concept biohybrid; with a central server, the Control Unit (CU), physically situated in Dresden, Germany; A neuromorphic (AiCTX, Zurich, Switzerland) and a biological (UNIPD, Padova, Italy) setup, plus a memristive setup SOTON (Southampton, UK)}
\label{fig_threeways_setup}
\end{figure}

The communication scheme introduced above (see section \ref{sec_UDP_AER}) was tested with four distributed setups forming one biohybrid. Fig.\ref{fig_threeways_setup} shows the setup used in the Synch project; a similar setup has been used in the Ramp project. The setup depicted in figure \ref{fig_threeways_setup} follows a star-shaped topology, in which a central server, the Control Unit (CU), physically situated in Dresden, Germany, monitors and routes firing activity from an neuromorphic (AiCTX, Zurich, Switzerland) and a biological (UNIPD, Padova, Italy) setup. The CU utilizes a third setup, SOTON (Southampton, UK), to implement a form of adaptation: Volatile Memristive Devices at SOTON are stimulated upon receiving events from both the artificial and biological neuronal populations via UDP. Their resistive state is thus reflecting an estimate of the inputs average firing frequency.  The Memristor Setup performs periodic read operations that are communicated back to the CU. Here, the source channel and the corresponding conductivity of the Memristor used are forming the contents of the UDP packet.  For this proof of concept, the neuromorphic setup (AiCTX) is configured as a simple spike repeater. In future, based on the frequency information from the memristive setup, long-term adaptation mechanisms can be implemented at the site of the CU, to provide control of the spiking neural networks firing dynamics in the neuromorphic setup.

\begin{figure}[h]
\centering
\includegraphics[width=1\textwidth]{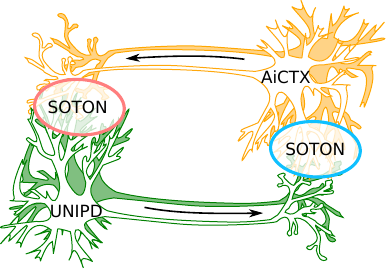}
\caption{Network equivalent of the biohybrid linkup}
\label{fig_threeways_setup_bio_circuit}
\end{figure}

Fig. \ref{fig_threeways_setup_bio_circuit} depicts the biological equivalent circuit of the biohybrid. In the communication from the biological population to the artificial one, Memristors can play a crucial role in pre-processing the event stream to reduce noise. This can be done by filtering the extracted firing frequency to suppress artefacts in the spike-detection and in the data-acquisition itself. The so processed firing rates serve as inputs to a Poisson-Process, that performs a conversion back into event-based communication and connects to the neurons of the Neuromorphic Processor.

Also in the communication towards the UNIPD electrophysiology setup, an expansion of the UDP protocol is beneficial: The Neurostimulation which forms a feed-back path from the artificial to the biological population uses parameters that go beyond the binary presence or absence of an incoming event. Indeed, traffic can be reduced by communicating the pulse-length and amplitude to be applied to the tissue, as an alternative to encoding this information in a pulse-frequency modulation.

\begin{table}
\scriptsize
    \centering
    \begin{tabular}{|l|c|c|c|c|} \hline 
         \textbf{Function}&  \textbf{Block 1}&  \textbf{Block 2}&  \textbf{Block 3}& \textbf{Block 4}\\ \hline \hline 
          \textbf{Raw traffic (Event-based)}&  &  &  & \\ \hline 
         Spike event from UNIPD via CU to& Setup ID & Timestamp & Reserved & Channel ID \\ 
        AiCTX&  &  &  & \\ \hline
        Stimulation event from AiCTX via CU & Setup ID & Timestamp & Weight & Neuron ID \\ 
        to UNIPD&  &  &  & \\ \hline
          &  &  &  & \\ \hline 
         \textbf{Write to Memristor (Event-Based)}&  &  &  & \\ \hline 
         Spike event from UNIPD via CU to  & Setup ID & Timestamp & Reserved & Channel ID \\ 
         SOTON &  &  &  & \\ \hline
         Spike event from AiCTX via CU to& Setup ID & Timestamp & Reserved & Neuron ID \\  
         SOTON&  &  &  & \\ \hline 
          &  &  &  & \\ \hline 
         \textbf{Read from Memristor (Periodic)}&  &  &  & \\ \hline 
         Artificial neurons global firing frequency& Setup ID & Reserved & Firing Rate & Reserved \\ 
         readout, from SOTON to CU&  &  &  & \\ \hline 
         Individual electrode channel firing & Setup ID & Reserved & Firing Rate & Channel ID \\ 
         frequency readout, to CU&  &  &  & \\ \hline 
         Update parameter from CU to & Setup ID & Parameter ID & Parameter MSB & Parameter LSB \\ 
         AiCTX&  &  &  & \\ \hline
         Initialization of parameters from CU to & Setup ID & Parameter ID & Parameter MSB & Parameter LSB \\ 
         AiCTX&  &  &  & \\ \hline
    \end{tabular}
    \caption{Packet structure for the individual communication modes used in the SYNCH network. Block 1 serves as an aid on how to interpret the subsequent blocks. All blocks are  64-bit in length}
    \label{tab:block_definitions}
\end{table}

All of the scenarios above have been addressed by implementing the following additions beyond event communication, see table \ref{tab:block_definitions}. Appended to every event package is a unique sender ID. This allows interpretation of an arriving package independent of the possibly changing sender IP addresses and instructs the receiver on how to process incoming data. In order to distinguish parameter information (e.g. weight updates) from events, the sender has a set of identifiers to choose from, in accordance with the receiver device capabilities.
Events coming from the Neuromorphic Processor are time-stamped in inter-spike intervals (ISI) and transmitted to the SOTON setup along with the identity of the source neuron. A unique identifier in the first Byte of the payload allows the SOTON setup to recognize the 24bit timestamp as ISI information.

At SOTON the system receives the events, translates the ISI timestamps into absolute time (T). The Neuromorphic Synapses then registers the spike as a pre-synaptic event, their resistive states are assessed and used according to the schemes and algorithms developed by SOTON and TUG. Subsequently, SOTON sends a series of parameter updates back to the Neuromorphic Processor, and the stimulating events forward to the UNIPD electrophysiology setup.

Stimulation can be communicated by sending: a) the SOTON identifier, b) the ID of the target synapse to stimulate, c) a measure of the weight (8 bits) located in the second reserved byte and d) the Timestamp of the “Post-Synaptic potential” (PSP) in absolute time T. UNIPD can then receive this information, and deliver stimulation where required. Importantly, UNIPD uses the weight to determine the intensity of stimulation on each target and the timestamp in order to set its own time reference. Any neuron that fires at the UNIPD set-up will be timestamped in absolute time as determined by the amount of `physical time' $(T + \Delta t)$ elapsed between the last synaptic PSP arriving at the said neuron and the time of firing. In other words if neuron N emits a spike 12 time steps after the last PSP (arriving at e.g. AT 15000), UNIPD will send an AER packet to SOTON informing the system that cell N has fired at AT 15012. The UNIPD response to SOTON thus allows to relate the firing of each post-synaptic neuron to an absolute time frame of reference and compute when and where plasticity should be implemented. 
In this way the described extension of the UDP protocol enables the compensation of transission-imposed delays in the time critical aspects of the system where Spike-Timing dependent Plasticity is performed. 
Furthermore uniquely identifying the senders is a method that can be extended for remote configuration and other uses in the future.

\begin{figure}[h]
\centering
\includegraphics[width=1\textwidth]{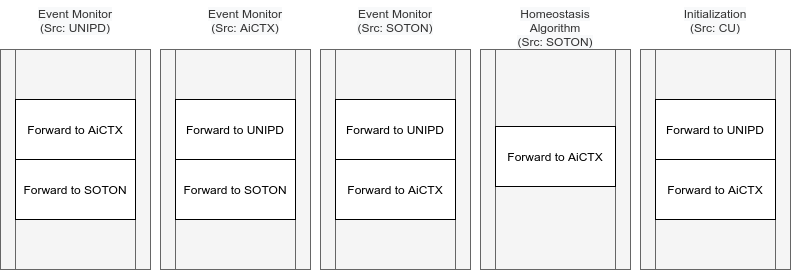}
\caption{Threads used in the UDP server that is part of the CU system}
\label{fig_udp_threads}
\end{figure}

As seen in Fig. \ref{fig_udp_threads}, the CU system is multi-threaded in order to process data incoming from all other setups and route it appropriately without the loss of packets and in order to support low-latency in the processing of the information received. Beside individual threads for the monitoring of incoming events from UNIPD, SOTON and AiCTX, an additional thread is opened in order to issue periodic parameter updates to the Neuromorphic Processor, implementing homeostasis and plasticity rules in the future. All threads open individual sockets for the communication via UDP, with assigned destination IPs and ports. This measure allows to conveniently log the firing activity and message passing for verification purposes.

\subsection{Plasticity/Timing requirements between setups}
\label{sec_plasticity_timing_requirements}

As outlined in Sec. \ref{sec_threeways_setup_intro}, an absolute timestamp can be introduced that allows at least locally accurate pre- and postsynaptic spike times, and via the central server, these local pre- and postsynaptic events can even be transmitted together so that a remote setup has the relative timings of e.g. its own sent events with the arrival time at the remote setup (forming the presynaptic events of the other setup) and the corresponding postsynaptic events at the other setup available. Thus, even spike-timing dependent  (STDP) processing paradigms are partially possible despite the delay. Other forms of processing or plasticity that are governed by e.g. rate coding are inherently robust to the delays introduced, thus present even less of a problem in a remote biohybrid.

As shown in \citep{henker2012accuracy}, even for very timing sensitive plasticity protocols like STDP, transmission jitter of up to 10~ms tends to still produce reasonably accurate network-level synaptic weight statistics (wrt an analytical/exact solution), as plasticity jitter tends to cancel out over several spike pairing. The plasticity/processing paradigm used in \citep{serb2020memristive} is partially rate-based, which further relaxes timing requirements. In fact, individual spikes in the rate code between memristor and bio setup are assumed to be from a time-varying Poisson distribution \citep{izhikevich2003relating}, so any jitter along the communication path just slightly modifies this distribution, but has no other macroscopic effect on the biohybrid. For the paradigm used in \citep{keren2018closed}, i.e. individual biological and neuromorphic networks connected in a row, with a feedback loop via a PID controller, the loop can tolerate about 100ms for the regulation to still be effective in steering the response probability of the individual neuron populations. 

\section{Results}
\subsection{results bidirectional link characterization}
\label{sec_results_bidirectional}

\begin{figure}[h]
\centering
\includegraphics[width=1\textwidth]{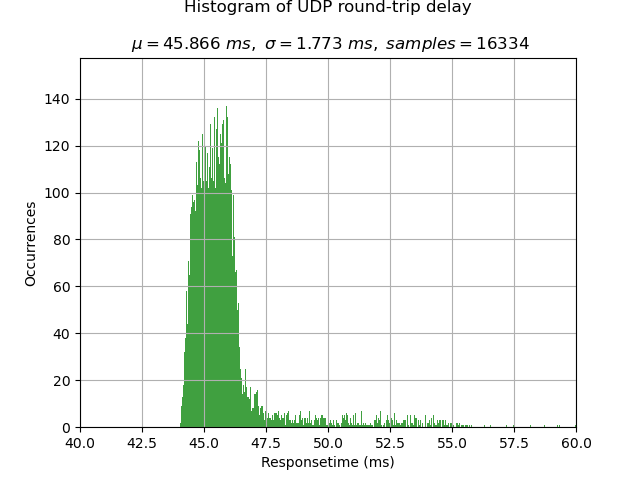}
\caption{Delay and Jitter of a UDP roundtrip}
\label{fig_roundtrip_UDP}
\end{figure}

Fig. \ref{fig_roundtrip_UDP} shows an experiment charcterizing an example of the connection pathways in Fig. \ref{fig_threeways_setup}, specifically the one between Padua and Dresden. The mean delay is 45.9~ms, jitter 1.77ms. Total packets dropped was 65. In addition, a further xx packets arrived after a delay of more than 5 seconds, so we count them in the packet loss, as compensating for this kind of delay would make the setups significantly too slow wrt each other. Thus, overall packet loss is xx/16334=0.398\%. In this particular sequence of packets, no corrupted packets were detected from CRC checksum. Also, no packets arriving in the wrong order were observed. In general, both those types of errors were in the single packet range over hour-long experiments, so negligible.

In terms of real-time processing respectively throughput, we also clocked our C++ software implementation (compare Fig. \ref{fig_udp_threads}).  we can get something like 20-40 kPulses (UDP packets) per second. This is in line with numbers in literature, with e.g. \cite{prytz2005real} reporting 25-30~$\mu$s for UDP rerouting in C++ under Linux. For the UDP core on an FPGA in Coronet, we easily achieved the 250kEvents/s required by the setup \citep{keren2018closed}. FPGA UDP implementations can go up to MEvents on e.g. a 1GBit/s Ethernet connection. Dedicated chip solution goes up to GEvents/s packet rates \citep{hartmann10}. 

Checking this against the five constraints defined in the beginning gives: (1) Packet loss: There is a significant fraction of packets either lost or so delayed so as not to meaningfully contribute to the biohybrid processing anymore, but still well below 1 in 200, which is on the lower end of pulse loses reported for AER-type communication in literature \citep{thanasoulis2014pulse}. As most processing and plasticity schemes contain some form of redundancy or averaging, this loss rate is acceptable. For constraint (2), i.e. real-time reception at the target system, the most relevant in our communication context are the delays due to encoding/decoding of packets. Even with a software-only implementation, this is on the order of 30-100$\mu s$, i.e. compatible with real-time processing. Constraint (3), i.e. transmission delay, lies somewhere between 30 and 60~ms on the internet connections we measured. While significant for some biohybrid paradigms, these can be partially compensated and/or are not so critical for most applications, as detailed in Sec. \ref{sec_plasticity_timing_requirements}. 
Regarding constraint 4, the 1-2ms jitter observed in our experiments can easily be compensated by delaying the packets by the max jitter and using the timestamps to align them. In most cases, even that is not required: As outlined in Sec. \ref{sec_plasticity_timing_requirements}, even plasticity schemes dependent on exact spike timing produce the right weight distribution for this small amout of jitter. Consequently, in the experiments documented in the next section respectively in \citep{serb2020memristive}, we did not compensate for jitter, rather letting ``time model itself'', i.e. spike packets are forwarded to the neuromorphic or bio setups as they come in. With respect to constraint (5) packet order: As stated above, virtually no packets arriving in the wrong order were observed. While compensation for out-of-sequence packets would have been possible via the timestamp, it was found easier to just discard these seldom occurences. 

\subsection{Results threeways biohybrid setup}
\label{sec_results_threeways}

\begin{figure}[h]
\centering
\includegraphics[width=1\textwidth]{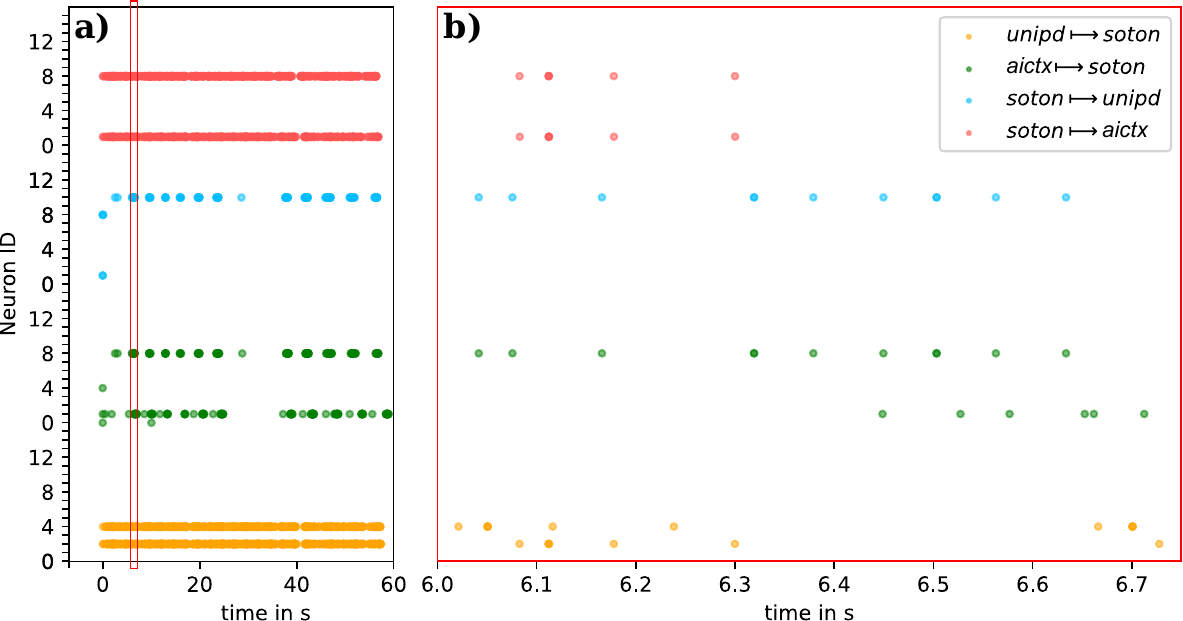}
\caption{Resulting spike trains from the proof-of-concept fourways biohybrid outlined in Fig. \ref{fig_threeways_setup}; (a): full recording across 60 seconds; (b) enlarged section of (a) for 6 to 6.75 seconds, showing individual spikes.}
\label{fig_threeways_setup_result}
\end{figure}

For the proof of concept shown in Fig. \ref{fig_threeways_setup} resp. Fig. \ref{fig_threeways_setup_result}, only two of the available 16 neuron recording channels have been activated; The signal chain begins with the UNIPD (cell culture) setup, the orange spike trace in Fig. \ref{fig_threeways_setup_result} (a) resp (b). These spikes are forwarded via the CU to the memristive setup at SOTON. The memristors are configured to integrate across the rate coming from the biological setup, providing a time-varying estimate of the biological fire rate. For details on this memristive integration, see \citep{serb2020memristive}. The Memristor Setup performs periodic read operations on the memristive states that are communicated back to the CU, i.e. the corresponding conductivity of the Memristor forms the contents of the udp packet. Based on this equivalent biological frequency information, gating based on firing rates,  thresholds for adaptation mechanisms or similar can be implemented at the site of the CU, e.g. to provide control of the artificial neurons spiking neural networks firing dynamics. Apart from this firing rate extraction, the SOTON setup forwards the biological spike traces from UNIPD to the neuromorphic setup at AICTX, denoted by the red spike traces (SOTON to AICTX) in Fig. \ref{fig_threeways_setup_result}. This serves as simple proof that the memristive setup can communicate with the neuromorphic setup. The AICTX setup uses these spikes from the biological setup as synaptic inputs for two neurons, which in response produce the two green spike traces. These are then downselected in the CU to one channel used for stimulation of the UNIPD setup. In future, the memristive setup at SOTON will take the function of synapses between the two populations (biological and circuit neurons), doing various preprocessing functions and passing the result to the neuromorphic setup, with the CU computing higher-level analysis and adaptations that act as configuration for the neuromorphic side. 

Please note that this paper is primarily about the UDP communication method for biohybrids. Thus, this experiment only serves as proof of concept for establishing a biohybrid remotely via UDP. More detailed biohybrd experiment results for the biohybrid of the Ramp project can be found in \citep{serb2020memristive}, for the Coronet project biohybrid in \citep{keren2018closed}. Publications detailing results for the Synch project biohybrid are in draft. Of course, these biohybrid experiments also use more of the transmission capacity that the UDP interface offers, not just two channels, and they employ more of the adaptation mechanisms possible in this four-ways setup, which are only hinted at above.

\section{Discussion}
As stated in the intro, we set out to design and implement a communication method that enables the interconnection of  complex, disparate and partially experimental technologies such as analog neuromorphics, biological cell cultures and memristors for a biohybrid \citep{tzouvadaki2023interfacing,mikhaylov2020neurohybrid}, from the safety of their home labs with their dedicated support infrastructure. In keeping with neuromorphic biohybrid information coding, this method should be event-based, similar to the original AER standard. 

For this, we implemented an UDP based, event/AER coded communication packet, based on some earlier work \citep{Rast2013}. The UDP standard helped to de-risk the combined setup significantly, as each setup could stay in its home lab, and only had to provide a standardized way of transmitting a UDP packet. The results section shows that the relevant characteristics (Packet loss, encoding/decoding delay at the host, transmission delay along communication pathway, jitter and packet order loss) are compatible with a real-time biohybrid or can be improved by e.g. our timestamp transmission to a point where they are fully in line with even e.g. spike time based learning \citep{george2020plasticity}. 

 \begin{figure}[!hbt]
     \centering
    \includegraphics[width=\textwidth]{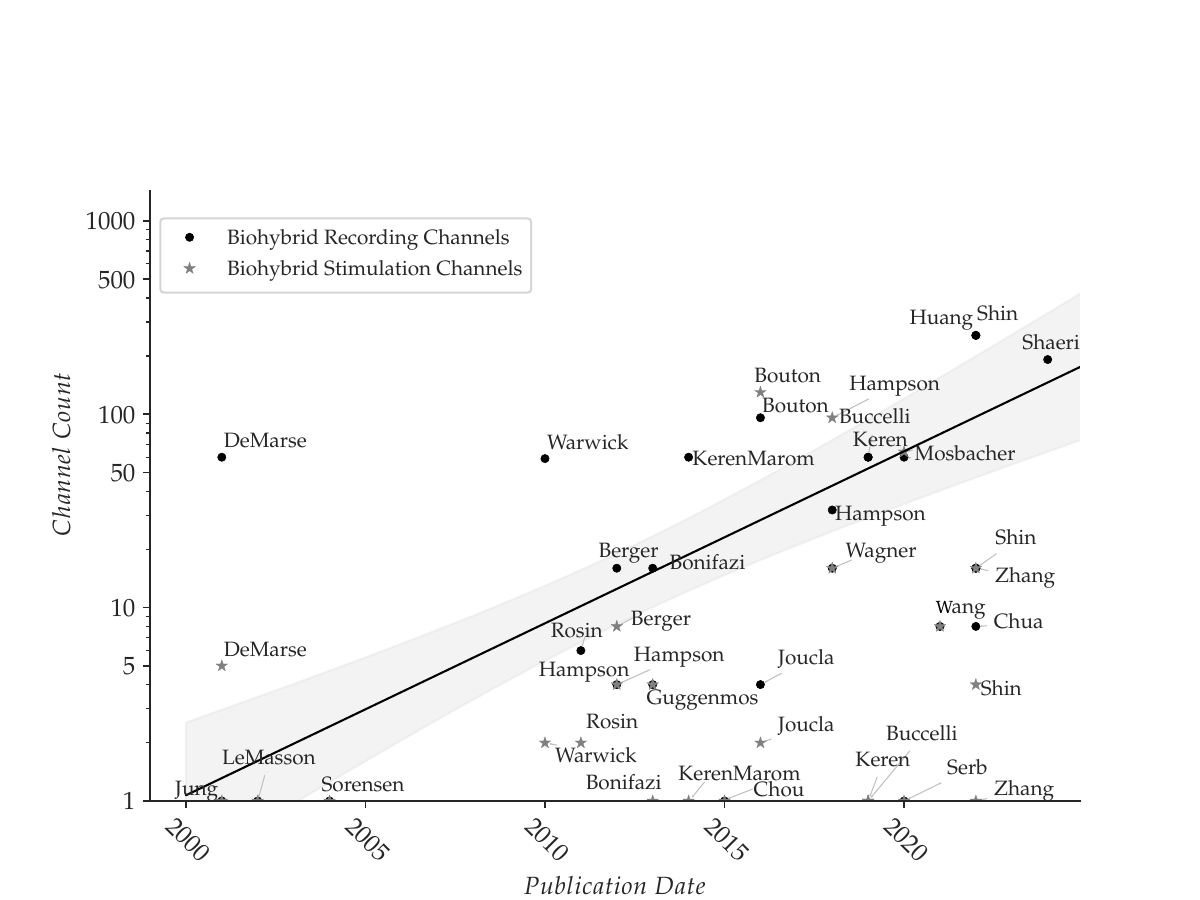}
     \caption{Advancements in the count of recording channels in biohybrid systems.}
     \label{fig:rec_stim_vs_time}
 \end{figure}

Encoders/decoders for the resulting UDP packet format have been implemented in various software platforms (Python, C++, Matlab) and in Verilog for FPGAs (or even ASIC realization). Obtainable data rates range from 20kEvents/sec up to MEvents/sec. In Fig. \ref{fig:rec_stim_vs_time}, we show the scaling of recording/stimulation channels in neuromorphic biohybrids over time. Assuming average events rates of up to 10 Hz per single channel, it is clear that even for future $>$1000 channel biohybrids, our remote biohybrid interconnect scheme can be used for the foreseeable future. 

We only showed a simple proof-of-concept of one concrete biohybrid setup in the results section, as this paper is more focussed on the method, not the biohybrid results. Biohybrid results using the UDP interconnect have e.g. been published for the Ramp project \citep{serb2020memristive} and the Coronet project \citep{keren2018closed}. Multiple papers for biohybrid results of the Synch project using the UDP interconnect are currently in draft. Specifically, in the Synch project, we link four setups across Europe, Israel and UK: one analog neuromorphic, one digital neuromorphic, one memristor, one neurobiology setup. This is starting to rival the most complex linkups achieved for local parallel AER, e.g. the Caviar setup \cite{serrano2009caviar}). 

Interestingly, for the Coronet biohybrid, we used the UDP setup first remotely to debug the overall biohybrid in the individual labs. In a second phase, the building blocks of the biohybrid were integrated locally in a lab at Technion Haifa, but still using the UDP interconnect, with everything connected to a local Ethernet switch. Even for a local setup, this method of interconnection has the advantage of simplicity compared to implementing parallel AER connections or other custom interfaces of e.g. the Microelectrode array, or the neuromorphic setup. On a local network, the relevant characteristics of course improved significantly, with e.g. no packet loss, an order of magnitude less timing jitter (compare \citep{bregni2009active}), and roundtrip delays below a millisecond, similar to measurements in literature \cite{prytz2005real}. Another benefit which we employed in the Coronet project is the ability to very simply inject background stimuli into the biohybrid by playing back an event file from a harddisk and sending it out via UDP.  

There are some examples in literature that have also used Ethernet packets to interconnect neurobiological signals, e.g. \cite{ierache2014navigation} sent EEG signals over the internet to remote-control a lego robot, or \citep{warwick2003application} connected a neuroprosthetic hand remotely to a patient. However, these look like one-of examples to demonstrate remote connection for the respective setups. Compared to these pure tech demonstrators, we think of our UDP connection scheme more as an enabler technology that accelerates biohybrid research by alleviating the need for a lot of travel between labs. This is also proven by the widespread use it has seen across three biohybrid EC projects, namely Coronet, Ramp and Synch. In those projects, the UDP interface was usually planned as a fallback, but has in all of them seen more use than the physical, local interconnect of the biohybrid components that was originally planned. Incidentally, having this remote setup also allowed us to continue biohybrid work throughout all the lockdowns and travel restrictions of the Covid pandemic, significantly speeding up our research. 

Besides it's use for biohybrids, versions of this UDP event coding have also been used to interconnect between neuromorphic setups, like SpiNNaker1 with the Brainscales wafer system \citep{Rast2013}, or SpiNNaker2 with automotive Radar processing inside a car\citep{vogginger2022automotive}. 
One of the current development efforts around SpiNNaker2 are remotely connected AI networks, where distributed versions of the event-based recurrent architectures of \citep{subramoney2022efficient} are processing data, e.g. for smart farming. The bidirectional stream of events for inference and learning \citep{subramoney2022efficient} between remote nodes will likely also be connected with this UDP framework. 

In a more neuroprosthetic or implant direction, \cite{chatterjee2023bioelectronic} propose an ``internet of bodies'' by remotely connecting bioelectronic sensor nodes. In a concrete example of this, \cite{pais2013brain} carry out an interesting experiment where one rat encodes a stimulus, and a second rat needs to take a decision based in microstimulation derived from the neural encoding of the first rat. The setup used in this experiment is only local and still fairly low dimensional, i.e. only a single scalar value gets transmitted. With our interface and the in-vivo systems available in Padua, we could move towards a significantly higher-dimensional version of this, as we support a direct spike-based interface. Compare also Coronet, which essentially was an in-vitro stand in for exactly this kind of whisker-based sensory decision task in rats. 

\bibliographystyle{plainnat}
\bibliography{biblio}

\end{document}